\documentclass[aps,prb,twocolumn,showpacs]{revtex4}
\usepackage{amsmath}
\usepackage{graphicx}
\usepackage[ansinew]{inputenc}
\usepackage{array}
\usepackage{color}
\usepackage{amsxtra}
\usepackage{amstext}
\usepackage{amssymb}
\usepackage{latexsym}
\usepackage{dsfont} 
\begin{document}

\title{How to detect level crossings without looking at the spectrum}

\author{M. Bhattacharya}
\affiliation{Department of Physics, University of Arizona, Tucson,
Arizona 85721}

\date{\today}
\begin{abstract}
We remind the reader that it is possible to tell if two or more
eigenvalues of a matrix are equal, without calculating the eigenvalues. We then
use this property to detect (avoided) crossings in the spectra of quantum
Hamiltonians representable by matrices. This approach provides a pedagogical
introduction to (avoided) crossings, is capable of handling realistic
Hamiltonians analytically, and offers a way to visualize
crossings which is sometimes superior to that provided by the spectrum. We illustrate the method using the
Breit-Rabi Hamiltonian to describe the hyperfine-Zeeman structure of the ground state
hydrogen atom in a uniform magnetic field.
\end{abstract}


\maketitle

\section{Introduction}
Crossings and avoided crossings occur in the spectra of many physical systems
such as atoms, \cite{walkup1998note} molecules, \cite{Yarkonyrmpnote} semiconductors 
\cite{sibille1998note} and microwave cavities. \cite{wiersig2006note} They can occur as a result
of tuning an external (e.g. electric) field \cite{Moruzzibook} or as the consequence of 
varying an internal (e.g. internuclear) coordinate, \cite{NWignersymm1929note} for instance. A number of
interesting physical phenomena have been associated with such (avoided) crossings.
For example an eigenstate transported in a closed circuit around a crossing in
the spectrum picks up a Berry phase. \cite{Berry1984note} As another example, points in the spectrum
where a large number of eigenvalues cross correspond to `hidden' symmetries of
the physical system. \cite{yuzbashyan2003note} These symmetries are hidden in the sense
that they are not evident \textit{a priori} as observables
that commute with the Hamiltonian. As a third example eigenvalue avoidance in
the spectrum signals the emergence of quantum chaos in a physical system. \cite{Haakebook}

Inspite of its importance as a basic phenomenon ubiquitous in physics, few
introductory texts treat (avoided) crossing in any
detail. Some questions that arise in the context of simple physical systems, and 
that could be considered at the (under)graduate level, are :

\begin{enumerate}
\item Is there a way to predict the total number of (avoided) crossings in the spectrum ?
\item Is there a systematic way to locate all the (avoided) crossings in the spectrum ?
\item Is there a way to identify the physical mechanisms responsible for
the occurrence of (avoided) crossings in the spectrum ?
\item Can the degeneracies - if any - in the spectrum be thought of as crossings ?
\item Sometimes the crossings are hard to discern in the spectrum - can they be visualized
in a clearer way ?
\end{enumerate}

In this article we employ an algebraic method that addresses these basic questions as
well as some others. It allows for a simple but systematic approach to
(avoided) crossings. In describing this approach we reintroduce to the study of (avoided) crossings a very
useful but perhaps neglected mathematical tool, the discriminant. 
\cite{Heiss1991note,Stepanov1998note,Davis2002note,freund2004note} In a series
of articles, \cite{mishprlnote,mishpra1note,mishpra2note} we have demonstrated in detail how use of algebraic
techniques is a powerful way to locate (avoided) crossings in the spectra of
quantum mechanical systems. We have shown that they are not only capable of locating (avoided) crossings 
without requiring solution of the Hamiltonian - a fact well known to mathematicians and not unknown
to physicists - they can also find (avoided) crossings when the Hamiltonian is not completely determined. \cite{mishpra1note} 
As one example, algebraic techniques allow us to derive a new
class of invariants of the Breit-Rabi Hamiltonian. These
invariants encode complete information about the parametric symmetries of the
Hamiltonian. \cite{mishpra2note} As another example, the use of algebraic methods 
allows us to detect the breakdown of the Born-Oppenheimer
approximation for molecules assuming only that the complicated molecular potentials are real. \cite{mishpra2note}

Although in the work just mentioned algebraic techniques were used in the context of
advanced research, namely Feshbach resonances, we show below that they also form 
an effective pedagogical tool. Indeed we believe the exposition in this article
could easily be included in the physics curriculum for (under)graduates as a striking demonstration
that enhances their understanding of quantum mechanics as well as linear algebra.

The rest of the paper is arranged as follows. Section II contains a simple
mathematical introduction using a 2$\times$2 matrix. Section III generalizes this to
the case of an $n\times n$ matrix . Section IV demonstrates the technique developed so far on the
ground state hydrogen atom in a uniform magnetic field. Section V
suggests some exercises for the reader; Section VI supplies a discussion.

\section{A 2 $\times$ 2 matrix}
\label{sec:two}
In order to motivate the general case we first consider a real-symmetric 2$\times$2 matrix
\begin{equation}
\label{eq:M2} H(P)= \begin{pmatrix}
        E_{1} &  V\\
        V & E_{2} \\
      \end{pmatrix},
\end{equation}
with (unknown) eigenvalues $\lambda_{1,2}$. The notation implies that all the
matrix elements depend on some tunable parameter $P$. Also $E_{1,2}$ could be the
bare energies of a two-level quantum system, which are mixed by the
perturbation $V$. Usually to find $\lambda_{1,2}$ we solve the equation
\begin{equation}
\label{eq:detM}
|H(P)-\lambda|=0,
\end{equation}
where $\lambda$ is a parameter. Eqs.(\ref{eq:M2}) and (\ref{eq:detM}) yield 
\begin{equation}
\label{eq:charpoly}
\lambda^{2}+C_{1}\lambda+C_{0}=0
\end{equation}
where $C_{0}= E_{1}E_{2}-V^{2}$ and $C_{1}=-(E_{1}+E_{2})$. However $\lambda_{1,2}$
also satisfy Eq.(\ref{eq:detM}), i.e.
\begin{equation}
\label{eq:charpoly2} (\lambda-\lambda_1)(\lambda-\lambda_2) = 0.
\end{equation}
Comparing the coefficients of Eqs.(\ref{eq:charpoly}) and (\ref{eq:charpoly2}), we find
\begin{eqnarray}
\label{eq:coefflambda}
\begin{array}{ll}
C_{0}=\lambda_{1}\lambda_{2},\\
C_{1}=-(\lambda_{1}+\lambda_{2}).\\
\end{array}
\end{eqnarray}
We are interested in crossings in the spectrum of $H(P)$ and therefore consider
the discriminant \cite{Laxbook} defined by
\begin{equation}
\label{eq:delta} 
\Delta \equiv (\lambda_{1}-\lambda_{2})^{2}.
\end{equation}
A little tinkering with Eqs.(\ref{eq:coefflambda}) shows that 
the discriminant can be re-written solely in terms of the coefficients of
Eq.(\ref{eq:charpoly}) :
\begin{equation}
\label{eq:delta2} \Delta = C_{1}^{2}-4 C_{0}.
\end{equation}
It is important to note that we did not actually calculate $\lambda_{1,2}$ at any
point in the discussion so far. Clearly, $\Delta = 0$ whenever a level crossing
occurs in the spectrum of $H(P)$. For example, if $E_{1}=E_{2}=P$, and $V=2P,$
then $\Delta =4P^{2}$, and there is a level crossing at $P=0$. This may be verified
by calculating the eigenvalues $\lambda_{1,2}=P,3P$. Note that a single crossing
in the spectrum corresponds to a double root of the discriminant.

We see from this example that use of the discriminant transforms the
problem of finding crossings in the spectrum to a polynomial root-finding
problem.  Further, it enables us to avoid calculating the eigenvalues. Lastly, it
provides the locations of all the crossings in the spectrum. In the next section
we generalize these statements to the case of an $n \times n$ matrix.

\section{An $n \times n$ matrix}
\label{sec:nn}
For a real symmetric $n \times n$ matrix $H(P)$ all of whose entries
are polynomials in $P$ the characteristic polynomial is
\begin{equation}
\label{eq:detn}
|H(P)-\lambda|=\sum_{m=0}^{n} C_{m}\lambda^{m},
\end{equation}
where the coefficients $C_{m}$ are all real. The discriminant is defined as \cite{Laxbook}
\begin{eqnarray}
\begin{array}{l}
\label{eq:Discdefn} 
D[H(P)]\equiv
\displaystyle\prod_{i<j}^{n}(\lambda_{i}-\lambda_{j})^{2},\\
\end{array}
\end{eqnarray}
in terms of the $n$ eigenvalues $\lambda_{i}$. It can also be written 
purely in terms of the $n+1$ coefficients $C_{m}$ : \cite{Basubook} 
\begin{equation}
\label{eq:Sylvester}  \small D[H(P)] =
\frac{(-1)^{\frac{n\left(n-1\right)}{2}}}{C_{n}}
\begin{vmatrix}
            C_{n}     & C_{n-1} & ...   & C_{0} &        &        &      \\
                      &  C_{n}  &       & ...   & C_{0}  &        &      \\
                      &         &       &       & ...    &        &      \\
                      &         &       &       &        & ...    & C_{0}\\
            nC_{n}    & ...     &       &       &        &        &      \\
                      & nC_{n}  & ...   & C_{1} &        &        &      \\
                      &         &       & ...   &        &        &      \\
                      &         &       &       &  .. .  & 2C_{2} & C_{1} \\
          \end{vmatrix}.
\end{equation}
In practice we will calculate discriminants of characteristic polynomials using the built-in 
{\fontfamily{pcr}\selectfont Discriminant} function in \textit{Mathematica} \cite{Mathnote} or \textit{Maple}. 
In addition we will use the following `toolbox' of general results in our discussion of (avoided) crossings below :
\begin{enumerate}
 \item The real roots of $D[H(P)]$ correspond, as in Section \ref{sec:two}, to crossings in
the spectrum of $H(P)$. 
 \item The real parts of complex roots of $D[H(P)]$ correspond to avoided crossings in the
spectrum of $H(P)$. For a proof, see Ref.\onlinecite{Heiss1991note}.
 \item A crossing is defined as the intersection of two eigenvalues. Hence the simultaneous 
intersection of \textit{m} eigenvalues gives rise to 
$\left( {\begin{array}{*{10}c} m \\ 2 \\ \end{array}} \right)=m(m-1)/2 $ crossings.
 \item Every (avoided) crossing contributes, as in Section \ref{sec:two}, a factor quadratic in $P$ to $D[H(P)]$.
       For a full proof, see Ref.\onlinecite{mishpra1note}.
 \item Since $H(P)$ is real symmetric, the eigenvalues $\lambda_{i}$ are all real. It follows from
Eq.(\ref{eq:Discdefn}) that $D[H(P)] \geq 0$. Hence Log$[D[H(P)]+1] \geq 0$ and goes to zero at every crossing. 
We will plot this function in order to visualize crossings.
\end{enumerate}
\section{The Hydrogen Atom}
\label{sec:H}
A simple but real example of a physical system whose spectrum exhibits both crossings
and avoided crossings is a ground state hydrogen atom in a uniform magnetic field. 
\cite{Dicksonnote} Such an atom is accurately described by the Breit-Rabi 
Hamiltonian \cite{breitrabi1931note}
\begin{equation}
\label{eq:BreitRabiHamiltonian} H_{BR}=A  \textbf{I}  \cdot
\textbf{S} + B (a S_{z}+ b I_{z}),
\end{equation}
where $I$ and $S$ indicate the proton and electron spin respectively,
and $B$ is the magnetic field along the $z$-axis. 
$A$ equals the hyperfine splitting and $a=g_{e}\mu_{B}$ and $b=g_{p}\mu_{N}$ where
$g_{e(p)}$ are the electron(proton) gyromagnetic ratios and $\mu_{B(N)}$ are the Bohr
(nuclear) magnetons respectively. The numerical values for these constants 
were obtained from Ref.\onlinecite{arimondormpnote} and are to be found in the figure captions in this article. 
In the basis $|M_{I},M_{S}\rangle $ which denotes the projections 
of $I$ and $S$ along $B$, the states are
\begin{equation}
 \label{eq:basis}
|\frac{1}{2},\frac{1}{2}\rangle,|\frac{-1}{2},\frac{1}{2}\rangle,
|\frac{1}{2},\frac{-1}{2}\rangle,|\frac{-1}{2},\frac{-1}{2}\rangle. 
\end{equation}
In this basis, the representation of Eq.(\ref{eq:BreitRabiHamiltonian}) is
\begin{equation}
\label{eq:HBRmatrix} H_{BR} = \tiny \frac{1}{4}
\begin{pmatrix}
  A-2(a+b)B     &                  &               &            \\
                & -A+2(a-b)B       & 2A            &            \\
                & 2A               & -A-2(a-b)B    &            \\ 
                &                  &               &  A+2(a+b)B \\
\end{pmatrix}.
\end{equation}
In order to calculate discriminants the basis (\ref{eq:basis}) is all we need but
we would also like to correlate our results to the spectrum, where we will use
the basis $|F,M_{F} \rangle$ to label the eigenstates of (\ref{eq:HBRmatrix}).
Here $F=I+S$ is the total angular momentum and $M_{F}$ its projection along the magnetic field.
This labelling is loose as strictly speaking in the presence of a magnetic field $M_{F}$ is
a good quantum number but $F$ is not.
\subsection{The $b=0$ case}
\label{subsec:bzero}
To begin we recollect that $|b|\ll |a|$, since the proton is 
more massive than the electron. \cite{arimondormpnote} We set $b=0$ in Eq.(\ref{eq:HBRmatrix}), 
calculate its characteristic polynomial, and find the discriminant to be
\begin{equation}
\label{eq:discnob}
D[H_{BR}^{b=0}]= \frac{1}{16}A^{4}a^{6}B^{6}\left(A^{2}+a^{2}B^{2} \right).
\end{equation}
Considering $D[H_{BR}^{b=0}]$ as a polynomial in $B$, the total number of its roots 
equals its degree, namely 8. Using the results 1., 2. and 4. from Section \ref{sec:nn} 
we can then say that the total number of the avoided crossings \textit{plus} 
the number of crossings in the spectrum of $H_{BR}^{b=0}$ is exactly 8/2=4, 
a fact verified below. Thus the discriminant allows us to predict the number of
(avoided) crossings before they are actually found.

We now systematically locate all the (avoided) crossings in the spectrum.
There is a 6-fold real root at $B=0$, which points to the only crossings in the spectrum (Section
\ref{sec:nn} 1.). Conventionally these
`zero-field' crossings are called \textit{degeneracies} and are not considered to be crossings. Hence we 
will maintain that there are \textit{no} crossings in the spectrum of $H_{BR}^{b=0}$. Eq.(\ref{eq:discnob}) has two complex roots:
\begin{equation}
\label{eq:Bcomplex}
B=\pm \frac{A}{a}i.
\end{equation}
This conjugate pair corresponds to a single avoided crossing at $B=0$ (Section \ref{sec:nn} 2. and 4.). 
From Eq.(\ref{eq:Bcomplex}) we can see that $A$, the hyperfine coupling of the two spins, is the physical mechanism 
responsible for the avoided crossing. Setting $A=0$ turns the avoided crossing into a zero-field
degeneracy.

\begin{figure}
\includegraphics[width=0.48 \textwidth]{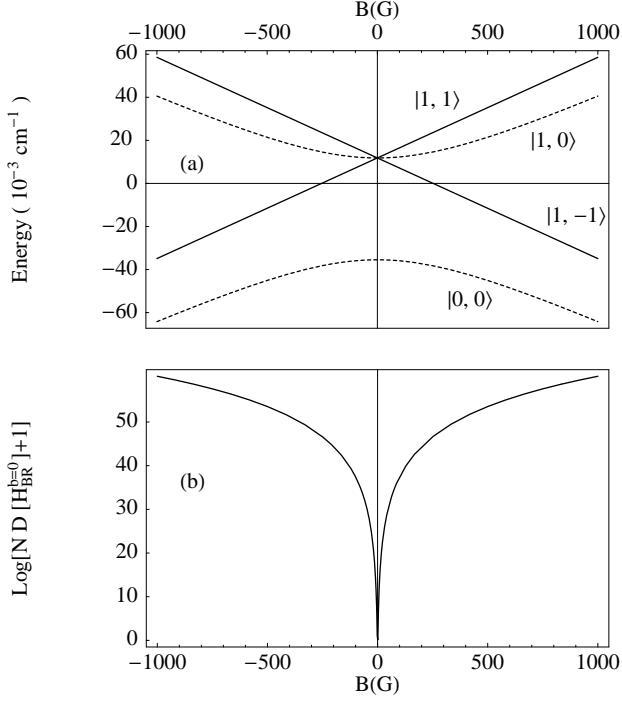}
\caption{\label{fig:HBRHnob} (a) The spectrum of $H_{BR}^{b=0}$ 
for atomic hydrogen ($S=1/2,I=1/2$) with states labelled using the $|F,M_{F}\rangle$ basis. 
The parameters used were $A=0.0473$cm$^{-1}$ 
and $a=9.35 \times 10^{-5}$cm$^{-1}$G$^{-1}$. The spectrum was calculated by analytically diagonalizing
the matrix in Eq.(\ref{eq:HBRmatrix}) and numerically plotting the eigenvalues. (b) Log$[N_{0} D[H_{BR}^{b=0}]+1]$
plotted using the same parameters as in (a) and the scaling factor $N_{0}=10^{35}$ used to optimize visibility.
The overall logarithmic trend is clearly visible.}
\end{figure}

We can confirm the predictions of the discriminant (Eq.\ref{eq:discnob}) by looking at 
Fig.\ref{fig:HBRHnob}(a). 
At $B=0$, 
three states in the upper manifold coincide, giving rise to 
$\left( {\begin{array}{*{3}c} 3 \\ 2 \\ \end{array}} \right)
=3$ crossings (Section \ref{sec:nn} 3.). Further, each crossing 
contributes a factor quadratic in $B$ (Section \ref{sec:nn} 4.)
to the discriminant (Eq.\ref{eq:discnob}), which therefore 
should - and does - contain a factor of $B^{6}$. This shows 
that the discriminant accounts for the (hyperfine) degeneracies 
of the spectrum by counting them as crossings.

An avoided crossing occurs at $B=0$ between the states drawn with dotted lines. In Fig.\ref{fig:HBRHnob}(b) a 
logarithmic representation of $D[H_{BR}^{b=0}]$ shows a dip corresponding to the degeneracies at $B=0$.
\subsection{The $b \neq 0$ case}
When $b \neq 0$, we find from Eq.(\ref{eq:HBRmatrix})
\begin{eqnarray}
\label{eq:discb}
\begin{array}{cl}
D[H_{BR}] = & \frac{1}{16}(a+b)^{2}B^{6}\left[A^{2}+(a-b)^{2}B^{2}\right]^{2}\\
            & \times \left[(a+b)A+2abB\right]^{2}\left[(a+b)A-2abB\right]^{2}.\\
\end{array}
\end{eqnarray}
The 6-fold real root at $B=0$ persists from the $b=0$ case (Eq.\ref{eq:discnob}), but now there are (`authentic') crossings for 
$B \neq 0$ at 
\begin{equation}
\label{eq:Bbcross}
B=\pm \frac{(a + b)A}{2ab}.
\end{equation}
`Switching' $b$ off and on therefore reveals the physical mechanism behind the appearance of crossings in the
Breit-Rabi spectrum : it is the interaction of the proton spin with the external magnetic field $B$.

The complex roots of Eq.(\ref{eq:discb}) are  
\begin{equation}
\label{eq:Bbav}
B=\pm \frac{A}{a-b}i
\end{equation}
and imply an avoided crossing at $B=0$ as in Section \ref{subsec:bzero} (Eq.\ref{eq:Bcomplex}).

\begin{figure}
\includegraphics[width=0.48 \textwidth]{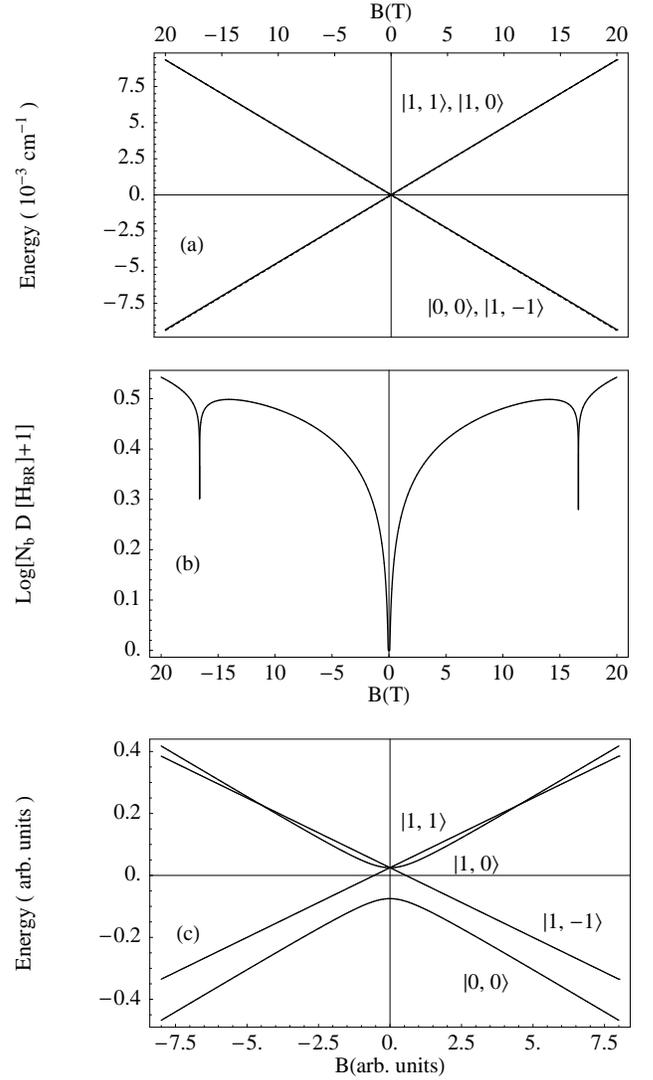}
\caption{\label{fig:HBRHb} (a) The spectrum of $H_{BR}$ 
for atomic hydrogen ($S=1/2,I=1/2$). The parameters used were $A=0.0473$cm$^{-1}$,
$a=9.35 \times 10^{-5}$cm$^{-1}$G$^{-1}$ and $b=-1.4202 \times 10^{-7}$cm$^{-1}$G$^{-1}$. 
The spectrum was calculated by analytically diagonalizing
the matrix in Eq.(\ref{eq:HBRmatrix}) and numerically plotting the eigenvalues.
The crossings at $B = \pm 16.6$ T cannot be resolved on this scale. (b) Log$[N_{b} D[H_{BR}]+1]$
plotted using the same parameters as in (a) and the scaling factor $N_{b}=10^{10}$. 
All three crossings are distinctly indicated by the minima of the discriminant; those corresponding to crossings
at $B = \pm 16$T actually do reach zero, but appear shorter here due to the resolution of the graphics.
The overall logarithmic trend of the plot is evident. (c) The same spectrum as in (a) but redrawn using the
parameters $A=0.1, a=0.01,$ and $b=-0.1$. The larger relative value of $b$ ensures that
the crossings at $B \neq 0$ can be seen in the plot.}
\end{figure}

We can confirm the predictions of the discriminant
(\ref{eq:discb}) by looking at Fig.\ref{fig:HBRHb}. 
The crossings which occur at $\pm 16.6$ T (from Eq.\ref{eq:Bbcross}) for actual values of the hydrogenic 
parameters $a,b,$ and $A$ cannot be seen in the spectrum [Fig.\ref{fig:HBRHb}(a)] as the energy level pairs are 
separated by less than the width of the lines used to plot them, a point 
made earlier in this journal (see Fig.2 in Ref.\onlinecite{Dicksonnote}). However the logarithmic representation 
[Fig.\ref{fig:HBRHb}(b)] of the discriminant on the same scale clearly shows dips at all the crossings. 
A scaled spectrum has been shown in Fig.\ref{fig:HBRHb}(c) using a much 
larger relative value of $b$ in order to display the crossings clearly. This example illustrates how the discriminant 
can sometimes prove superior to the spectrum in displaying crossings. For another example see Ref.\onlinecite{mishpra2note}.
\section{Suggested Exercises}
\begin{enumerate}
 \item Prove that the product in Eq.(\ref{eq:Discdefn}) contains $n(n-1)/2$ factors.
 \item Justify the presence of $A^{4}$ in Eq.(\ref{eq:discnob}).
 \item For the parametric symmetry $a=b$ in Eq.(\ref{eq:discb}) show that there are crossings 
but no avoided crossings in the spectrum of $H_{BR}$.
 \item Plot $D[H_{BR}]$ as a function of $B$ and try to identify the zeros that correspond
to crossings. This is an exercise designed to show that the highly nonlinear nature
of the discriminant implies that each of its terms dominates in a different regime 
of $B$. This makes it difficult to include all the features of the discriminant on a single scale
unless a smoother representation - such as a logarithmic one - is adopted. 
 \item The Wigner von-Neumann non-crossing rule \cite{NWignersymm1929note} says 
``States of the same symmetry (i.e. quantum number) do not cross,
 except accidentally.'' Verify the rule for the $|M_{F}\rangle$ states in 
Figs.\ref{fig:HBRHnob}(a) and \ref{fig:HBRHb}(c). That is, show that
 states which (avoid) cross possess (same) different $M_{F}$'s.
\end{enumerate}
\section{Discussion}
The examples presented above illustrate that the discriminant is an elegant but simple device
for locating and counting (avoided) crossings. Further, it is an effective tool for investigating
the physical mechanisms behind the occurrence of (avoided) crossings. Lastly, visualization of
the discriminant offers an alternative to locating crossings in the spectrum. It must be noted 
however that the discriminant yields no information about which eigenvalues 
avoid or intersect, or about the eigenvectors. Also shallow avoided crossings do not show up in the
logarithmic representation, especially if they are near to crossings, which give rise to strong features in the
discriminant. For all such information the spectrum has to be consulted. 

The technique we have presented
can be used algorithmically on Hamiltonians which are polynomial in some parameter $P$, and which can be represented by 
finite dimensional matrices. Examples are a spin 1/2 particle in a magnetic field \cite{Merzbacherbook} (the archetypal two-level
system) and the hydrogen atom in an electric field \cite{Merzbacherbook}
(usually presented as an example of degenerate perturbation theory).
However the method can also yield insight into physical systems whose Hamiltonians are usually truncated 
to a finite dimension for practical calculations such as the nucleus modeled as a triaxial rotor, \cite{wood2004note} 
and a polar molecule in an electric field. \cite{Brownbook} Another interesting application is the calculation
of critical parameters of quantum systems, \cite{kais2006note} since the critical point 
occurs at a crossing. An example
using the Yukawa Hamiltonian has been treated in Ref.\onlinecite{serra1998note}.
A list of physical systems in atomic and molecular physics to which algebraic methods can be applied is provided in Ref.\onlinecite{mishpra2note}. 

To conclude we have presented an algebraic technique
for systematically analysing (avoided) crossings in the spectra of 
physical systems and pointed out its usefulness as a pedagogical device. 
It is a pleasure to thank P. Meystre for support. This work is supported in part by the US Army Research Office, 
NASA, the National Science Foundation and the US Office of Naval Research.

\end{document}